\documentclass[preprint]{aastex}
\usepackage{mkfig}

\begin{document}

\title{\bf Studying the Pulsation of Mira Variables in the Ultraviolet}

\author{B. E. Wood\altaffilmark{1}, M. Karovska}
\affil{Harvard-Smithsonian Center for Astrophysics, 60 Garden St., Cambridge,
  MA 02138.}
\email{wood@head-cfa.harvard.edu, karovska@head-cfa.harvard.edu}

\altaffiltext{1}
  {Present address: JILA, University of Colorado, Boulder, CO 80309-0440.}

\begin{abstract}

     We present results from an empirical study of the Mg~II h \& k emission
lines of selected Mira variable stars, using spectra from the
{\em International Ultraviolet Explorer} (IUE).  The stars all exhibit similar
Mg~II behavior during the course of their pulsation cycles.  The Mg~II flux
always peaks after optical maximum near pulsation phase $\phi=0.2-0.5$,
although the Mg~II flux can vary greatly from one cycle to the next.
The lines are highly blueshifted, with the magnitude of the blueshift
decreasing with phase.  The widths of the Mg~II lines are also
phase-dependent, decreasing from about 70 km~s$^{-1}$ to 40 km~s$^{-1}$
between $\phi=0.2$ and $\phi=0.6$.  We also study other UV
emission lines apparent in the IUE spectra, most of them Fe~II lines.
These lines are much narrower and not nearly as blueshifted as
the Mg~II lines.  They exhibit the same phase-dependent flux behavior as
Mg~II, but they do not show similar velocity or width variations.

\end{abstract}


\section{Introduction}

     At some point in their lives, many if not most stars go through an
unstable phase which leads to pulsation.  There are many classes of these
pulsating stars.  Perhaps the most famous are the Cepheid variables, which
are popular mostly because of their well-defined relationship between
stellar luminosity and pulsation period (typically 5--50 days) that
makes these stars very useful as distance indicators.

     Mira variables are another important class of stellar pulsators, having
long periods of 150--500 days and luminosities that vary by as
much as 6--7 magnitudes from minimum to maximum.  Miras are asymptotic
giant branch (AGB) stars with masses similar to that of the Sun.
They have very massive, slow, cool winds, which produce a complex
circumstellar environment.  Observations of molecular CO lines show that
Miras are often surrounded by molecular envelopes thousands of AU in
diameter.  These observations yield estimates of the wind termination
velocity and total mass loss rate, which are typically of order 5 km~s$^{-1}$
and $10^{-7}$ M$_{\odot}$ yr$^{-1}$, respectively \citep{ky95}.
The circumstellar envelopes are rich sites for dust formation and are often
found to be sources of SiO, OH, and H$_{2}$O maser emission
\citep{pjd94,bl97}.

     The massive winds of Miras are believed to be driven by a combination
of shocks induced by stellar pulsation, and dust formation
\citep{ghb88}.  The shocks lift a substantial amount of
material up to 1--2 stellar radii above the surface of the star.  Radiation
pressure on dust formed in this material then pushes it away from the star.
The pulsation-induced shocks not only assist in generating the massive winds
of Miras, but they also determine the atmospheric structure of these stars
to a large extent.  Thus, understanding the nature of the shocks and
measuring their properties is essential to understanding the physics of
pulsation and mass loss from pulsating stars.

     The ultraviolet spectral regime is an ideal place to study radiation
from the shocks.  Many UV emission lines are generated from
immediately behind the shocks, which are potentially very useful diagnostics
for various characteristics of the shocks.  Foremost among these lines
are the strong Mg~II h \& k lines at 2800 \AA.

     A large number of UV spectra of Miras have been taken by the
{\it International Ultraviolet Explorer} (IUE) over the years, and
some of the basic characteristics of the Mg~II h \& k lines have been
noted.  It is known, for example, that the Mg~II lines are not visible
throughout part of the pulsation cycle.  They typically appear at about
pulsation phase $\phi=0.1$, well after optical maximum ($\phi=0.0$).  The
Mg~II fluxes peak around $\phi=0.3-0.45$ and then decrease until becoming
undetectable at about $\phi=0.7$ \citep{ewb86,dgl96}.

     For a set of LW-HI observations of S~Car and R~Car, \citet{jab89}
showed that the Mg~II h \& k lines are blueshifted
relative to the stellar rest frame by as much as 100 km~s$^{-1}$, and the h
line is significantly stronger than the k line.  Both of these properties
are very difficult to explain, as the shock speeds should be much lower than
100 km~s$^{-1}$, and for other astronomical targets the k line is almost
always found to be stronger than the h line \citep[e.g.][]{rdr95}.

     Clearly the unusual behavior of the Mg~II lines of Miras
should be looked at in more detail to understand
the pulsation process.  In this paper, we utilize the extensive IUE
data sets that exist for 5 Miras to fully characterize the behavior
of the ultraviolet emission of these stars.  Many pulsation cycles are
sampled for each star, allowing us to see how the UV emission lines behave
from one cycle to the next.

\section{The IUE Archival Data}

     We have searched the IUE database for Miras that have been extensively
observed by the satellite.  In this paper, we are only interested in
Miras without companion stars that may be contaminating the UV spectrum.
We are particularly interested in high resolution spectra taken with
IUE's long-wavelength cameras (i.e.\ LW-HI spectra).  With these spectra,
fully resolved profiles of detected emission lines (especially Mg~II) can be
analyzed.  However, we also include low resolution, long-wavelength
camera spectra (i.e.\ LW-LO spectra) in our analysis, which can at least be
used to measure accurate fluxes of the Mg~II lines and the background.
We only consider large aperture data to ensure that our measured fluxes are
accurate.  Except for the Lyman-$\alpha$ line, which is blended with
geocoronal emission, no emission lines are typically seen in Mira spectra
taken with IUE's short-wavelength SWP camera, so we are not interested in
those data.

     Table 1 lists the five Miras with the most extensive sets of IUE
observations.  (Actually, L$^{2}$~Pup is a semi-regular variable star
rather than a Mira, but it is a long-period pulsating star similar to Miras
and it has a very large IUE dataset so we include it in our analysis.)  The
table gives the position, distance, center-of-mass radial velocity ($V_{rad}$),
and pulsation period of each star.  The distances were taken from the
Hipparcos catalog \citep{macp97}.

     The center-of-mass radial velocities of these stars were taken from the
SIMBAD database. However, deriving true systemic center-of-mass velocities is
difficult for the pulsating Miras, because different spectral features
observed for these stars exhibit different velocities, and these velocities
are often found to vary during the pulsation cycle.  The velocities in
Table 1 are based on measurements of optical absorption lines of neutral
atoms, which provide plausible values for $V_{rad}$.  Other potential
measures, such as various optical emission lines, molecular radio emission
lines, and maser lines, are generally blueshifted relative to the neutral
atomic absorption lines by roughly 5--10 km~s$^{-1}$.  These features are
probably formed either behind an outward moving shock or in the outer regions
of the stellar atmosphere where the massive wind of the star is being
accelerated \citep{ahj54,pwm60,gw75,dh88}.  Note that infrared CO emission
lines, which can also be used to estimate stellar center-of-mass radial
velocities, generally suggest velocities that are blueshifted relative to the
optical velocities by $\sim 5$ km~s$^{-1}$ \citep{khh78,khh96,khh97}.

     In order to determine how UV emission varies during the pulsation cycle
it is necessary to derive accurate pulsation phases, which requires knowledge
of the pulsation period and a zero-phase date.  All of the Miras in
Table 1 have been monitored for many decades, and their average periods are
well known.  However, the periods of Miras can sometimes differ from this
average.  Therefore, in order to derive the most accurate phases, we sought
to determine the average period and zero-phase of our Miras during IUE's
lifetime only (1978--1996).

     The American Association of Variable Star Observers (AAVSO) has a
long-term program to monitor hundreds of Miras using observations from
amateur astronomers around the world \citep{jam97}.
Using AAVSO data obtained from the World Wide Web, we derived the pulsation
periods listed in Table 1.  The derived periods of S~Car, R~Car, and R~Leo
are within a day of the accepted long-term average values
\citep[see, e.g.,][]{ah85}, but the derived period of
L$^{2}$~Pup is 4 days shorter and that of T~Cep is 11 days longer.  For these
two stars, the accuracy of the computed pulsation phases is improved
significantly with the use of the derived periods in Table 1.  As mentioned
above, L$^{2}$~Pup is a semi-regular variable rather than a traditional Mira,
but it is similar to Miras in other respects so it is included in our sample.
Because of the irregularity of its period, the accuracy of the pulsation
phases we derive is limited to about $\pm 0.2$.  The pulsation of the
other stars on our list is far more regular, and we estimate that the phases
we use for these stars are accurate to within about $\pm 0.05$.

     Table 1 lists the number of LW-HI and LW-LO
spectra available for each star.  In each case, the available data provide
reasonably good coverage of at least 2--3 different pulsation cycles.
The data were extracted from the IUE Final Archive.  The spectra in
the archive were processed using the NEWSIPS software, which became available
near the end of the IUE's 18-year lifetime in the mid-1990s.  This software
corrects the fixed pattern noise problem that plagued spectra processed with
the older IUESIPS software, and improves signal-to-noise by up to a factor
of 2 for high resolution spectra.  The NEWSIPS software is described in
detail by \citet{jsn96}.

\section{Data Analysis}

\subsection{The LW-LO Spectra}

     Figure 1 illustrates the behavior of the near UV spectrum of the Mira
R~Car during a typical pulsation cycle in 1986--1987.  The primary spectral
feature visible in the LW-LO spectra is the Mg~II h \& k feature at 2800~\AA,
which appears shortly after optical maximum, achieves a maximum flux around
$\phi=0.4-0.5$, and then declines.  The Fe~II UV1 multiplet lines at
2620~\AA\ are also visible when Mg~II is at its brightest.

     We developed a semi-automated routine to measure the Mg~II fluxes from
the LW-LO spectra of our stars.  For each spectrum, the underlying background
flux is estimated using the flux in the surrounding spectral region.
The reasonableness of this computed background is verified visually.  After
subtracting the background from the spectrum, the Mg~II flux is then computed
by direct integration.  The error vector provided by the NEWSIPS software
then allows us to estimate the 1~$\sigma$ uncertainty for this flux.  Three
spectra from R~Leo show substantial background emission at an unexpected
phase, which we believe indicates contamination by scattered solar light, so
these spectra are not used.

     In Figure 2, we plot the Mg~II fluxes as a function of pulsation phase
for all the stars in our sample.  We use thick symbols to indicate Mg~II
lines with 3 or more pixels that have been flagged by NEWSIPS as being
potentially inaccurate, usually due to overexposure.  These fluxes
should be considered more uncertain than the plotted 1~$\sigma$ error bars
would suggest.  In general, however, these fluxes are not wildly discrepant
from fluxes of better exposed lines measured closely in time.  Dashed lines
connect points that are within the same pulsation cycle, excluding cases
where the points are separated by over 0.4 phase.

     Figure 2 shows that the pulsation cycles of each star are all consistent
with the general rule of a flux maximum around $\phi=0.2-0.5$.  However,
there are substantial flux differences from one cycle to the next.  The R~Car
data set shows the most extreme such variability, with the Mg~II fluxes
reaching only about $1\times 10^{-13}$ ergs cm$^{-2}$ s$^{-1}$ in one cycle
and approaching $1\times 10^{-10}$ ergs cm$^{-2}$ s$^{-1}$ during another.
In contrast, the Mg~II fluxes of S~Car differ by less than an order of
magnitude from one cycle to the next, this despite the fact that S~Car has
one of the largest IUE LW-LO databases.  The other relatively short period
pulsator in our sample, L$^{2}$~Pup, seems to behave similarly.  The data
points show more scatter for L$^{2}$~Pup, but this is probably due to
inaccuracies in the estimated pulsation phases caused by the irregular
period of this star (see \S 2).  A comparison of L$^{2}$~Pup's Mg~II
and optical variability has been presented by \citet{ewb90}.

     The phase of maximum Mg~II flux differs somewhat for the five
stars, with S~Car appearing to have the earliest maximum and R~Leo the
latest.  One of the brighter Mg~II cycles of R~Leo is particularly
interesting for showing substantial Mg~II flux very near optical maximum at
$\phi=0.04$ (observations LWP21679 and LWP21680).  We generally do not detect
significant Mg~II flux this close to $\phi=0.0$ for any of our stars
(excluding L$^{2}$~Pup because of its uncertain phases), but apparently there
can be exceptions to this rule.  By inspecting the AAVSO light curve we
confirmed that the strong Mg~II lines had in fact been observed very near
optical maximum.

\subsection{The LW-HI Spectra}

\subsubsection{The Mg~II h \& k Lines}

     Turning our attention to the LW-HI spectra, in Figures 3--7 we display
the observed Mg~II h \& k line profiles for the five stars in our sample.
The k line is shown as a solid line and the h line is represented by a
dotted line.  The spectra are shown on a velocity scale in the stellar rest
frame, assuming the stellar center-of-mass velocities listed in Table 1, and
assuming rest wavelengths in air of 2802.705~\AA\ and 2795.528~\AA\ for
the h and k lines, respectively.  The date, time, and pulsation phase of
each observation are indicated in the figures.  Cosmic ray hits, which are
apparent in the original spectra as very narrow, bright spectral features
not observed in other spectra, have been identified and removed manually.
One S~Car spectrum with wildly discrepant properties (LWP12197) was removed
from the analysis and is therefore not shown in Figure 5.

     Many of the S~Car spectra were taken with the star deliberately offset
within IUE's large aperture.  This results in an inaccurate wavelength
calibration.  Fortunately, there is a set of nine observations taken on 1991
March 9--13 with different offsets (see Fig.\ 5).  After measuring the
centroids of the Mg~II h line (see below), we were able to see from these
data how the offsets affected the h line centroid and we used this
information to correct the wavelengths of all the S~Car spectra obtained
with aperture offsets.  The spectra in Figure 5 are the corrected spectra.

     In almost every spectrum in Figures 3--7, the k line is contaminated by
substantial absorption centered at a stellar rest frame velocity of about
$-70$ km~s$^{-1}$.  This absorption is due to lines of Fe~I and Mn~I with
rest wavelengths in air of 2795.005~\AA\ and 2794.816~\AA, respectively
\citep{dgl89,dgl98}.
Because the mutilation of the k line by these overlying neutral
absorbers is generally severe, we focus most of our attention on the more
pristine h line.

     Another concern is interstellar absorption, which should affect both
the h and k lines at the same velocity.  The one star whose spectrum we know
will {\em not} be affected by the ISM is S~Car, because its extremely large
$+289$ km~s$^{-1}$ center-of-mass velocity will shift the Mg~II lines well
away from any ISM absorption.  The h line of L$^{2}$~Pup shows absorption at
about $-40$ km~s$^{-1}$ (see Fig.\ 7).  The ISM velocity predicted for this
line of sight by the local ISM flow vector of \citet{rl95}
is $-43$ km~s$^{-1}$ after shifting to the stellar
rest frame.  Thus, it seems likely that this is indeed interstellar
absorption, although for lines of sight as long as those towards the stars
in our sample, ISM components could potentially be present with
significantly different velocities than that predicted by the local flow
vector.

     The h lines of R~Leo, R~Car, and T~Cep do not appear to be contaminated
by any obvious ISM absorption features.  For R~Leo and T~Cep, the ISM
velocities predicted by the local flow vector ($-4$ and $+14$ km~s$^{-1}$,
respectively) suggest that the ISM absorption may lie outside of the observed
Mg~II emission (see Figs.\ 3 and 6).  For the R~Car line of sight,
the local flow vector predicts ISM absorption at $-30$ km~s$^{-1}$, which
falls just within the red side of the h line.  No obvious absorption is seen,
however, suggesting that the ISM column density for this particular line of
sight may be quite low.  This is not necessarily unusual, as other lines of
sight with distances of $\gtrsim 100$ pc have been found to have very low H~I
column densities of $\sim 10^{18}$ cm$^{-2}$ \citep{cg95,np97}.

     We will assume that the Mg~II h line profiles of R~Leo, R~Car, and T~Cep
are at most only mildly affected by ISM absorption, meaning that for these
three stars and S~Car we can be reasonably confident that any substantial
line profile differences we might find are due to intrinsic differences in
the stellar spectra.  The Mg~II h line profiles of these four stars often
appear to be slightly asymmetric, with a red side that is steeper than the
blue side.  Nevertheless, the profiles can be represented reasonably well
as a Gaussian, and can therefore be quantified with Gaussian fits.
We developed a semi-automated procedure to fit all Mg~II h lines
detected with enough signal-to-noise (S/N) for a meaningful fit to be
performed.  The quality of all fits was verified visually.  For spectra
without strong Mg~II h lines, we estimated a line
flux by direct integration as we did for the LW-LO spectra.  For
L$^{2}$~Pup, we measured h line fluxes for {\em all} the spectra in this
manner, since we could not accurately fit Gaussians to the ISM-contaminated 
line profiles observed for this star.

     In Figures 8--10, we plot versus pulsation phase the Mg~II h line
fluxes, centroid velocities, and widths measured from the LW-HI spectra of
the stars in our sample.  As in Figure 2, dotted lines connect points within
the same pulsation cycle, and thick data points indicate h lines with three
or more pixels flagged as being overexposed.  The 1~$\sigma$ uncertainties
shown in the figure were estimated using the procedures outlined by
\citet{ddl92}.  The flux variations seen in Figure 8
are essentially the same as those seen in the LW-LO data (see Fig.\ 2).
Note once again how much the Mg~II fluxes vary from one cycle to the next
for R~Car.

     \citet*{jab89} found that the Mg~II lines in
selected observations of S~Car and R~Car were substantially blueshifted.
Figure 9 demonstrates that this behavior is common to all the Mg~II lines
observed by IUE for all the Miras in our sample.  Furthermore, Figure 9 also
reveals a strong correlation between line velocity and pulsation phase, in
which the magnitude of the line blueshifts decreases with pulsation phase.
For example, between $\phi=0.2$ and $\phi=0.6$ the line velocities of S~Car
change from about $-100$ km~s$^{-1}$ to $-50$ km~s$^{-1}$, and those of
T~Cep change from $-50$ km~s$^{-1}$ to $-30$ km~s$^{-1}$.  Note that even
though we are using Gaussians to fit the Mg~II profiles, this does {\em not}
imply that Mg~II h \& k are optically thin or that the large blueshifts of
these lines actually represent gas velocities (see \S 3.2.2).

     This velocity behavior exactly mimics the behavior of the optical Ca~II
H \& K lines of Miras, which also have blueshifted velocities that typically
change from about $-100$ km~s$^{-1}$ to $-40$ km~s$^{-1}$ between $\phi=0.2$
and $\phi=0.6$ \citep{pwm52,wb52}.  Thus, opacity effects and atmospheric
flow fields appear to induce the same line behavior in both the Mg~II and
Ca~II lines.  Furthermore, many Cepheid variables appear to exhibit
similar behavior, suggesting that these phase-dependent velocity variations
may be typical for stellar pulsators in general \citep{pwm60}.

     The data in Figure 9 suggest a possible correlation between pulsation
period and line velocity, with the shortest period Mira in our sample (S~Car)
having the largest blueshifts, and the longest period star (T~Cep) having the
smallest.  The R~Car data suggest another possible correlation.
For the pulsation cycle with the largest Mg~II fluxes, the Mg~II h lines of
R~Car are more blueshifted than they are during the
weaker cycles.  Unfortunately, our sample of stars and the number of
well-sampled pulsation cycles per star is very small, making it difficult to
truly establish these correlations.

     Figure 10 demonstrates that a well-defined phase dependence also exists
for the line widths, which are quantified in Figure 10 as
full-widths-at-half-maxima (FWHM).  The line width behavior is very similar
for all the stars, with a decrease from about 70 km~s$^{-1}$ to 40 km~s$^{-1}$
between $\phi=0.2$ and $\phi=0.6$.

\subsubsection{Other Lines in the IUE LW-HI Spectra}

     The Mg~II h \& k lines are by far the brightest lines that appear in
the LW-HI spectra of Miras, but they are not the only lines present.  During
pulsation cycles that produce strong Mg~II lines, other lines
also appear out of the background noise of the LW-HI spectra.

     The largest Mg~II fluxes observed for any star in our sample were
observed during the 1989--90 pulsation cycle of R~Car, during which Mg~II h
line fluxes reached up to $5\times 10^{-11}$ ergs cm$^{-2}$ s$^{-1}$ (see
Figs.\ 4 and 8).  Figure 11 shows three sections of an LW-HI spectrum taken
during this period.  These sections contain all of the obvious real
emission features apparent in the full spectrum.  The expected line locations
of several multiplets are indicated in the figure, which accounts for most of
the observed emission features.  The spectrum is dominated by several
multiplets of Fe~II lines (UV1, UV32, UV60, UV62, and UV63).  The
Mg~II h \& k (UV1) lines are of course apparent, as perhaps are two much
weaker Mg~II lines of the UV3 multiplet.  The intersystem Al~II] (UV1) line
at 2669.155~\AA\ is easily visible.  \citet{dgl98}
observed the feature at 2823~\AA\ in an HST/GHRS
spectrum of the Mira R~Hya, and identified it as an Fe~I (UV44) 2823.276~\AA\
line which is fluoresced by the Mg~II k line.  A couple other emission
features in Figure 11 can also be identified with Fe~I (UV44) lines.

     The lines seen in Figure 11 have previously been
observed in many IUE and HST/GHRS spectra of red giant stars
\citep{pgj91,kgc95,rdr98}.
The Al~II] line and most of the Fe~II lines are formed by collisional
excitation, as are the Mg~II h \& k lines, but \citet{pgj92}
find that the UV62 and UV63 multiplet lines of Fe~II
are excited by a combination of collisional excitation and photoexcitation
by photospheric emission at optical wavelengths.

     The flux behavior of all the lines in Figure 11 parallels that of Mg~II,
increasing to a maximum near $\phi=0.3-0.4$ and then decreasing.  Thus, the
Fe~II and Al~II] lines presumably originate in atmospheric locations similar
to Mg~II h \& k.  Note that many of the lines apparent in Figure 11 were
only observed during this one particularly bright cycle of R~Car, presumably
being too weak to be detected during weaker cycles on R~Car and the other
stars.  For a selection of the brightest emission lines, we used an automated
fitting procedure like that described in \S 3.2.1 to fit Gaussians to the
lines of all the stars in our sample, whenever the lines are clearly
detected.  The lines measured in this manner are listed in Table 2.
For L$^{2}$~Pup, no observed cycle was bright enough to detect any of these
lines, and for S~Car only the brightest Fe~II line at 2625.667~\AA\ could
ever be clearly detected.

     For the observations in which the lines are observed, we find that
line flux ratios among the various Fe~II lines are always similar to those
seen in Figure 11, as are Mg~II/Fe~II flux ratios.  For example, the
(Mg~II $\lambda$2803)/(Fe~II $\lambda$2626) ratio is always about 20.
The Fe~II and Mg~II/Fe~II flux ratios found for the Miras are somewhat
different from those observed in red giants, but not radically so.

     However, the Fe~II line {\em profiles} are very different.
For normal red giant stars, most of the Fe~II lines are clearly
opacity broadened and have profiles very different from that of a simple
Gaussian \citep*{pgj91,kgc95}.  In contrast, the Fe~II lines of
Miras are very narrow, Gaussian-shaped emission features, with widths at or
near the IUE's instrumental resolution of $\sim 0.2$ \AA.  This is why we
could accurately fit the lines with single Gaussians.  The exception is the
broader Fe~II 2599.394~\AA\ line, which should have the highest opacity of
all the Fe~II lines.

     The widths and velocities of the lines listed in Table 2 do not
appear to exhibit any substantial phase dependent behavior, in contrast to
Mg~II h \& k.  As an example, in Figure 12 we plot the velocities of the
Fe~II 2625.667~\AA\ line as a function of pulsation phase for the four Miras
in which this line was occasionally detected.  Only S~Car has a hint
of phase dependence, with some evidence for a small increase in velocity
with phase.  Much of the scatter seen in Figure 12 could be due to
uncertainties in target centering, which can induce systematic velocity
errors of $\pm5$ km~s$^{-1}$ \citep{bew95}.

     Since there is little if any phase dependence in the velocities of the
lines listed in Table 2, we compute a weighted average and standard deviation
for all the measurements of all the lines \citep{prb92}, and in Table 2
these velocities are listed for each Mira in
our sample.  For R~Leo, R~Car, and S~Car, all of the lines are blueshifted
$5-15$ km~s$^{-1}$.  The Fe~II 2599.394~\AA\ line is once again an exception,
showing significantly larger blueshifts.

     The lines of T~Cep behave differently, as they do not show systematic
blueshifts relative to the star.  The Mg~II lines of T~Cep are also not as
blueshifted as the other stars (see \S 3.2.1 and Fig.\ 9).  It is difficult
to identify a reason for this difference in behavior, but
perhaps the center-of-mass velocity we are assuming for this star is off by
$\sim 10$ km~s$^{-1}$.  The difficulties in defining center-of-mass
velocities for Miras have already been discussed in \S 2.

     The large blueshifts observed for the Mg~II lines are unlikely to be
direct measurements of outflow velocities.  The shock velocities present
in Miras are expected to be of order $10-20$ km~s$^{-1}$, although this issue
is still a matter of debate \citep*{ghb88,dh88,khh97}.  Thus, the
$5-15$ km~s$^{-1}$ blueshifts seen for most of the Fe~II and Al~II] lines of
R~Leo, R~Car, and S~Car are more likely to be measuring the true outflow
velocities of shocked material.  The larger widths and blueshifts of the
Mg~II lines are probably due to opacity effects, similar to the findings of
\citet{pgj93} for non-Mira M-type giants.  The similar behavior of the
Fe~II $\lambda$2599 line suggests that this line is influenced by similar
opacity effects, which is reasonable since this line should have the highest
opacity of any of the Fe~II lines.

     The other Fe~II lines will have substantially lower
optical depths than Mg~II h \& k and Fe~II $\lambda$2599, and some may even
be optically thin.  The Al~II] $\lambda$2669 line should also have very low
opacity, since it is a semi-forbidden transition.  Thus, the Fe~II and
Al~II] line centroids should be more indicative of the true outflow
velocities of the shocked material, and the line widths (which are actually
unresolved) should be more indicative of turbulent velocities within the
shocked material.

     Interpreting the difference between the behavior of the Mg~II lines and
that of the less opaque lines could be very important for understanding the
structure of the shocks propagating through Mira atmospheres.  In
future work, we hope to explore possible reasons why the Mg~II lines are
broader and more blueshifted than the less optically thick lines.

\section{Summary}

     We have compiled IUE observations of 5 Mira variables with substantial
IUE data sets in order to study the properties of emission lines seen in the
UV spectra of these stars, which are believed to be formed behind outwardly
propagating shocks in the atmospheres of these pulsating stars.  Our
findings are summarized as follows:
\begin{description}
\item[1.] We confirm the phase-dependent Mg~II flux behavior previously
  reported for Mira variables \citep*[e.g.][]{ewb86},
  which is observed for all the pulsation cycles that we study:  the Mg~II
  flux rises after optical maximum, peaks near $\phi=0.2-0.5$, and then
  decreases.  For some Miras (e.g.\ R~Car) the amount of Mg~II flux produced
  during a pulsation cycle can vary by 2--3 orders of magnitude from one
  cycle to the next, while for others (e.g.\ S~Car) the flux behavior is
  more consistent.
\item[2.] The Mg~II k lines are almost always contaminated with circumstellar
  absorption lines of Fe~I and Mn~I, making analysis of the line profile very
  difficult.
\item[3.] The Mg~II h line is always blueshifted, with the magnitude of the
  blueshift decreasing with pulsation phase.  The blueshifts vary somewhat
  from star to star and cycle to cycle, but typical velocity changes are from
  $-70$ km~s$^{-1}$ to $-40$ km~s$^{-1}$ from $\phi=0.2$ to $\phi=0.6$.  Note,
  however, that these line shifts do not represent the actual gas velocities
  at the formation depths of these lines, because of the high opacity of
  Mg~II h \& k.  These velocity variations are very similar to those of the
  optical Ca~II H \& K lines.
\item[4.] The width of the Mg~II h line decreases from about 70 km~s$^{-1}$
  to 40 km~s$^{-1}$ between $\phi=0.2$ and $\phi=0.6$.
\item[5.] In addition to the Mg~II lines, other lines of Fe~II, Fe~I, and
  Al~II] are also observed in IUE LW-HI spectra.  The fluxes of these lines
  show the same phase-dependent behavior as the Mg~II lines.
\item[6.] Unlike Mg~II, these other emission lines tend to be very narrow
  and do not show phase-dependent velocity and width variations.  Except
  for Fe~II $\lambda$2599, the Fe~II and Al~II] lines of most of the Miras
  show blueshifts of $5-15$ km~s$^{-1}$, which may indicate the flow velocity
  of the shocked material.  In contrast, the lines of T~Cep do not show any
  significant line shifts, although we speculate that perhaps this
  is due to an uncertain center-of-mass velocity for this star.
\end{description}

\acknowledgments

     We would like to thank the referee, Dr.\ D.\ Luttermoser, for many
useful comments on the manuscript.  MK is a member of the Chandra X-ray
Center, which is operated under contract NAS-839073, and is partially
supported by NASA.  In this research, we have used, and acknowledge with
thanks, data from the AAVSO International Database, based on observations
submitted to the AAVSO by variable star observers worldwide.

\clearpage

\begin{deluxetable}{lccccccc}
\tablecaption{Mira Variables with Large IUE Data Sets}
\tablecolumns{8}
\tablehead{
  \colhead{Star} & \colhead{RA} & \colhead{DEC} & \colhead{Distance} &
    \colhead{$V_{rad}$} & \colhead{Period} & \multicolumn{2}{c}{\# of
    IUE Spectra} \\
  \cline{7-8} \\
 \colhead{} & \colhead{(J2000)} & \colhead{(J2000)} & \colhead{(pc)} &
    \colhead{(km s$^{-1}$)} & \colhead{(days)} & \colhead{(LW-HI)} &
    \colhead{(LW-LO)}}
\startdata
S Car       &10:09:22 &$-61^{\circ}32^{\prime}57^{\prime\prime}$ &
  $405\pm103$ &+289 & 150 & 42 & 67 \\
R Car       & 9:32:15 &$-62^{\circ}47^{\prime}20^{\prime\prime}$ &
  $128\pm14$  & +28 & 308 & 28 & 45 \\
L$^{2}$ Pup & 7:13:32 &$-44^{\circ}38^{\prime}39^{\prime\prime}$ &
  $61\pm5$    & +53 & 136 & 15 & 68 \\
T Cep       &21:09:32 & $68^{\circ}29^{\prime}27^{\prime\prime}$ &
  $210\pm33$  &$-12$& 399 & 15 & 32 \\
R Leo       & 9:47:33 & $11^{\circ}25^{\prime}44^{\prime\prime}$ &
  $101\pm21$  & +13 & 313 & 13 & 58 \\
\enddata
\end{deluxetable}

\clearpage

\begin{deluxetable}{lcccccc}
\small
\tablecaption{Line Velocities}
\tablecolumns{7}
\tablehead{
  \colhead{Ion} & \colhead{Wavelength} & \colhead{Multiplet} & 
    \multicolumn{4}{c}{Velocity (km s$^{-1}$)} \\
  \cline{4-7} \\
 \colhead{} & \colhead{} & \colhead{} & \colhead{R Leo} &
    \colhead{R Car} & \colhead{S Car} & \colhead{T Cep}}
\startdata
Al II]&2669.155&1&$-7.1\pm8.7$  & $-4.7\pm3.3$ & \nodata     & $1.5\pm4.7$ \\
Fe I &2823.276&44&$-5.6\pm9.6$  & $-3.9\pm5.0$ & \nodata     & $3.2\pm4.0$ \\
Fe II&2598.368&1 &$-14.5\pm9.1$ & $-10.7\pm1.4$& \nodata     & $3.8\pm8.7$ \\
Fe II&2599.394&1 &$-24.8\pm4.8$ & $-36.1\pm2.6$& \nodata     &$-16.3\pm5.2$ \\
Fe II&2607.085&1 &$-14.4\pm5.9$ & $-10.0\pm2.8$& \nodata     &  \nodata \\
Fe II&2611.873&1 &$-12.9\pm2.5$ & $-8.5\pm2.1$ & \nodata     &$-3.8\pm10.0$ \\
Fe II&2617.616&1 &$-15.9\pm6.1$ & $-9.6\pm2.8$ & \nodata     &$-11.5\pm3.3$ \\
Fe II&2620.408&1 &$-9.5\pm8.4$  & $-5.9\pm2.7$ & \nodata     & $4.8\pm2.6$ \\
Fe II&2625.667&1 &$-12.8\pm6.0$ & $-10.7\pm2.7$&$-15.9\pm4.9$& $-2.7\pm1.8$ \\
Fe II&2628.292&1 &$-15.6\pm6.5$ & $-9.2\pm2.5$ & \nodata     & $-3.7\pm3.4$ \\
Fe II&2732.440&32&$-13.2\pm4.8$ & $-3.5\pm1.8$ & \nodata     & $3.0\pm3.7$ \\
Fe II&2759.334&32&$-9.1\pm4.3$  & $-5.6\pm1.9$ & \nodata     & $3.2\pm3.6$ \\
Fe II&2926.587&60&$-12.7\pm6.8$ & $-8.2\pm4.4$ & \nodata     & $1.5\pm4.2$ \\
Fe II&2953.773&60&$-9.0\pm3.5$  & $-4.1\pm3.8$ & \nodata     & $-2.0\pm4.8$ \\
Fe II&2730.734&62&$-18.2\pm10.4$& $-6.9\pm3.8$ & \nodata     &  \nodata \\
Fe II&2743.197&62&$-5.2\pm6.4$  & $-4.6\pm3.4$ & \nodata     & $6.2\pm4.1$ \\
Fe II&2755.735&62&$-9.3\pm6.3$  & $-8.1\pm2.6$ & \nodata     & $0.6\pm4.2$ \\
Fe II&2727.538&63&$-18.6\pm9.6$ & $-14.3\pm4.7$& \nodata     &  \nodata \\
Fe II&2739.547&63&$-12.8\pm8.3$ & $-7.2\pm3.4$ & \nodata     & $5.2\pm2.7$ \\
\enddata
\end{deluxetable}

\clearpage

\begin{figure}
\plotone{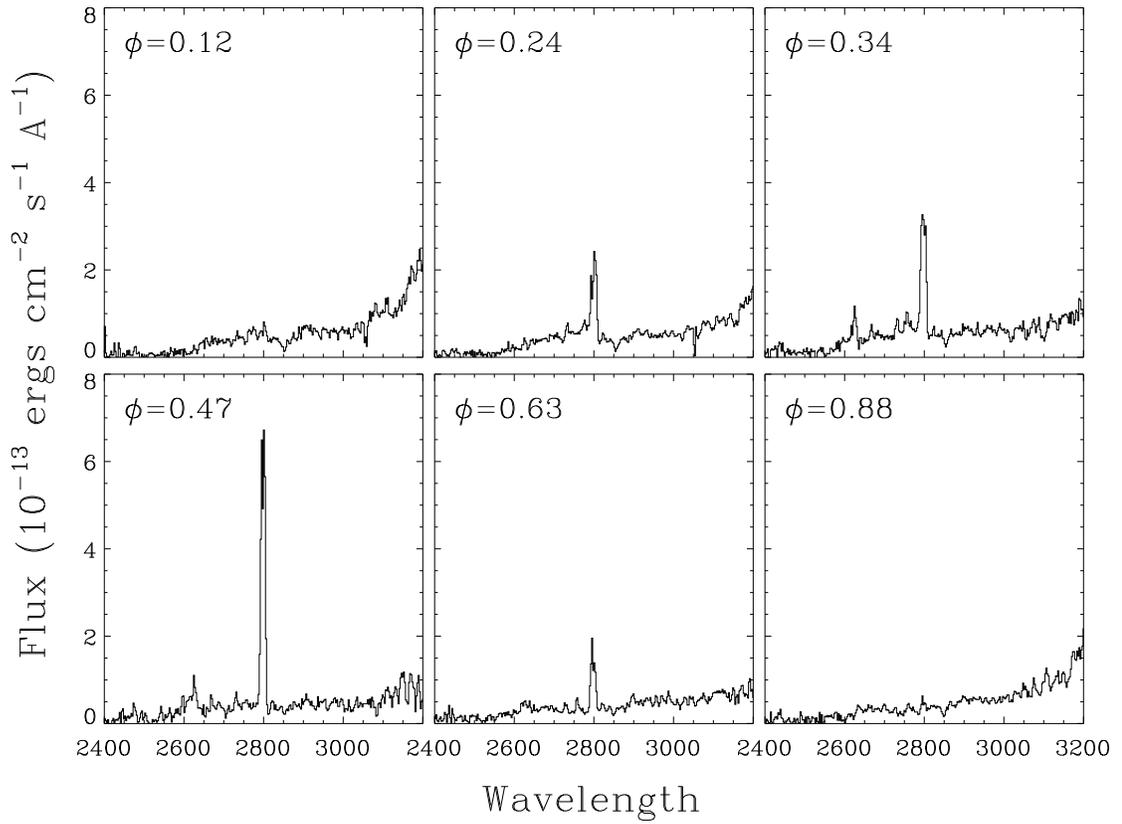}
\caption{Six low resolution IUE spectra of R~Car, illustrating
  the variation in Mg~II h \& k line flux as a function of pulsation phase
  for one pulsation cycle.}
\end{figure}

\clearpage

\begin{figure}
\plotfiddle{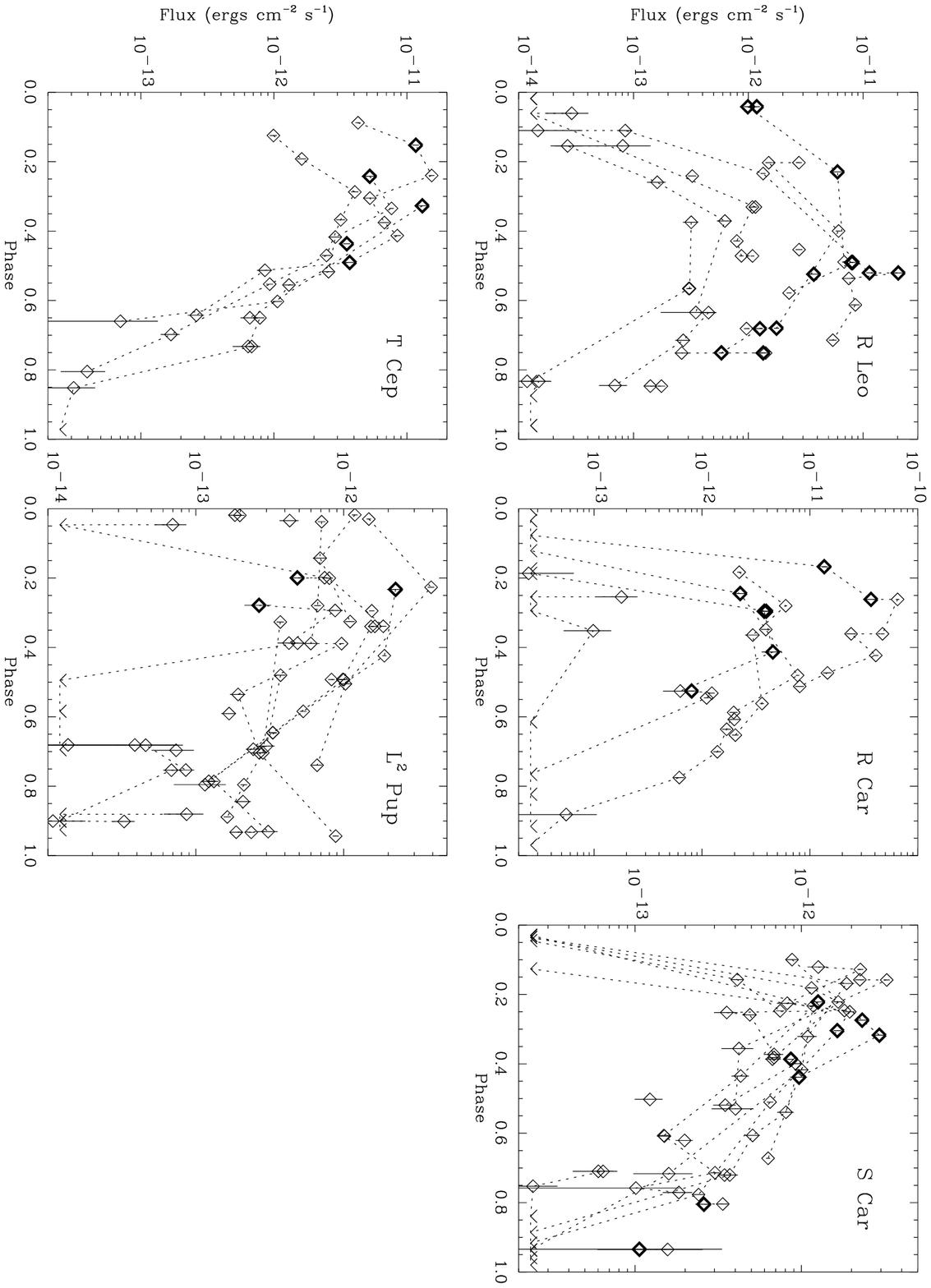}{3.5in}{90}{75}{75}{300}{0}
\caption{The Mg~II line fluxes measured from IUE LW-LO spectra, plotted
  versus pulsation phase.  Dotted lines connect points within the same
  pulsation cycle.  Thick symbols identify potentially inaccurate data
  points (usually due to overexposure), as indicated by NEWSIPS data quality
  flags.}
\end{figure}

\clearpage

\begin{figure}
\plotfiddle{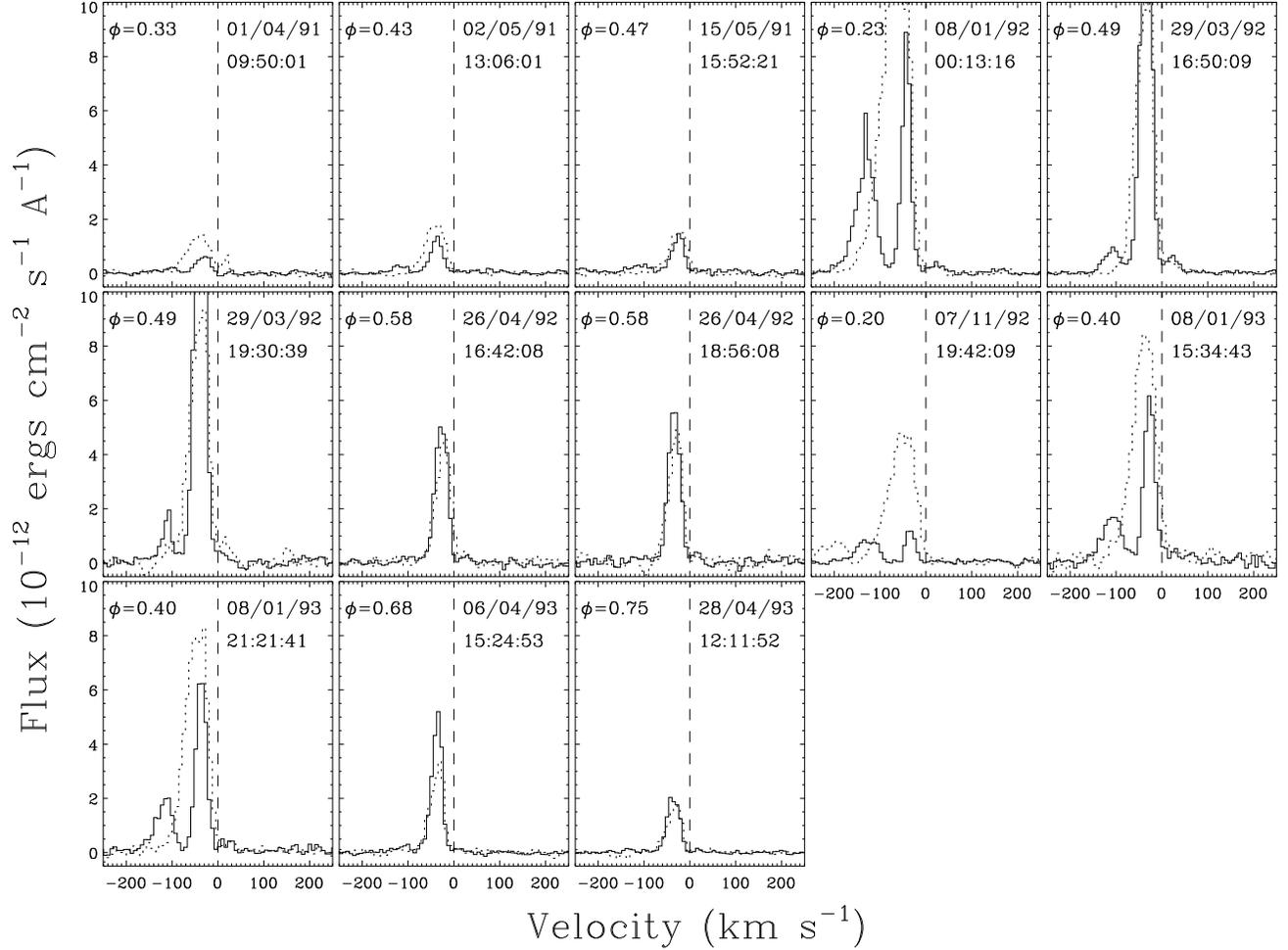}{3.5in}{90}{75}{75}{300}{0}
\caption{The IUE LW-HI observations of the Mg~II h (dotted lines) and k
  (solid lines) line profiles of R~Leo.  The pulsation phase, date, and UT
  time of observation are indicated in each panel.  The spectra
  are plotted on a velocity scale in the stellar rest frame.}
\end{figure}

\clearpage

\begin{figure}
\plotfiddle{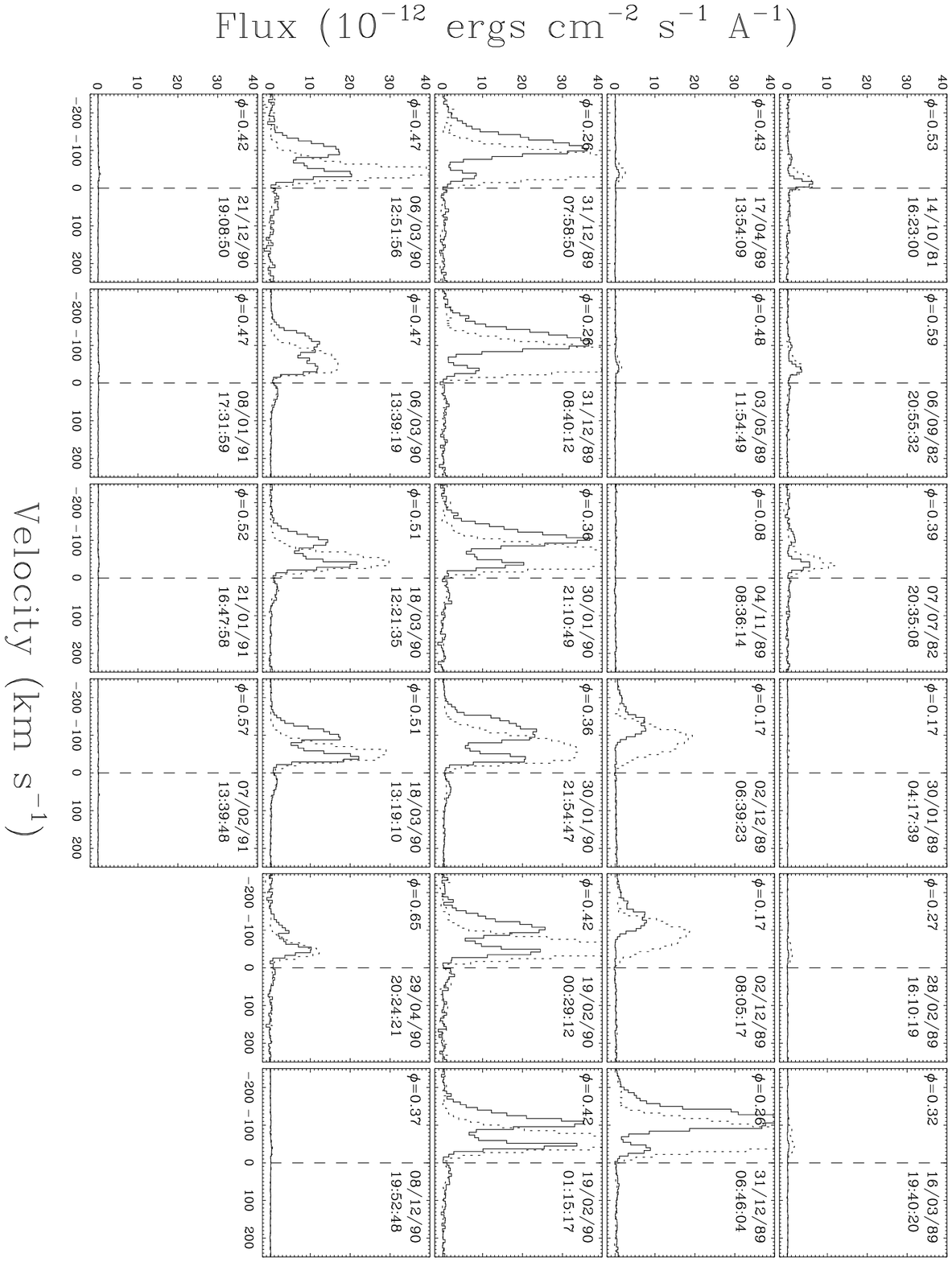}{3.5in}{90}{75}{75}{300}{0}
\caption{Same as Fig.\ 3, for R~Car.}
\end{figure}

\clearpage

\begin{figure}
\plotfiddle{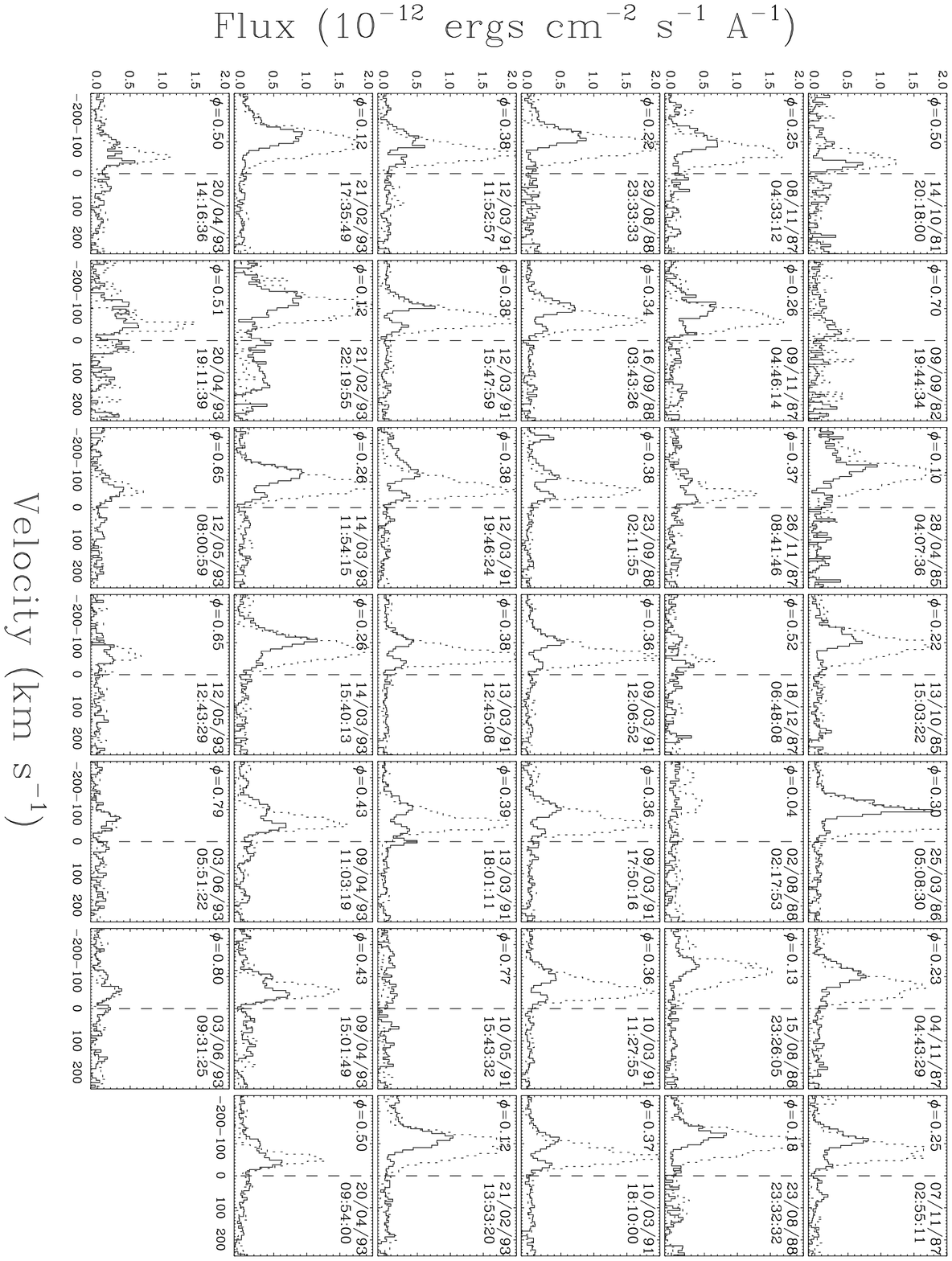}{3.5in}{90}{75}{75}{300}{0}
\caption{Same as Fig.\ 3, for S~Car.}
\end{figure}

\clearpage

\begin{figure}
\plotfiddle{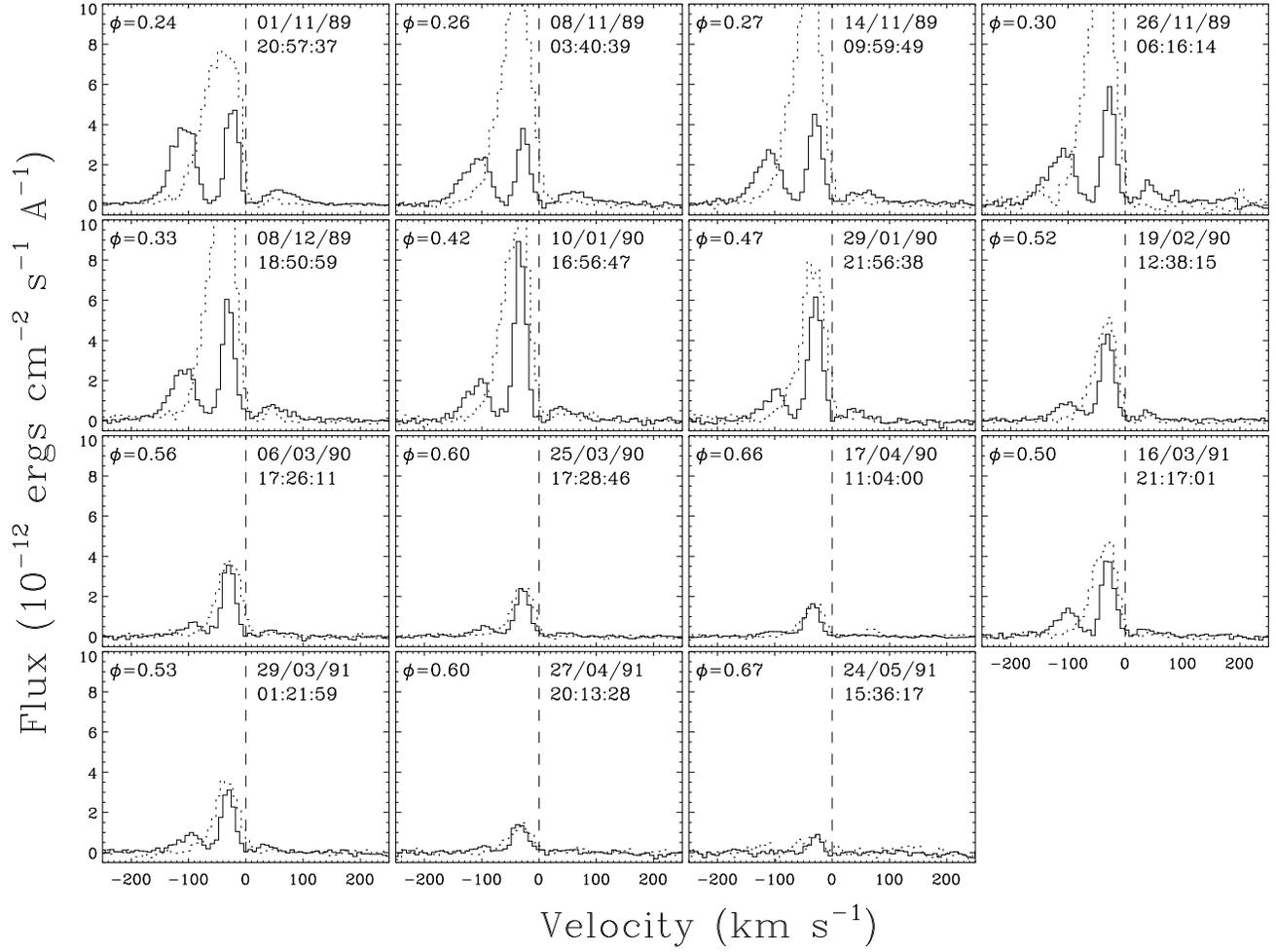}{3.5in}{90}{75}{75}{300}{0}
\caption{Same as Fig.\ 3, for T~Cep.}
\end{figure}

\clearpage

\begin{figure}
\plotfiddle{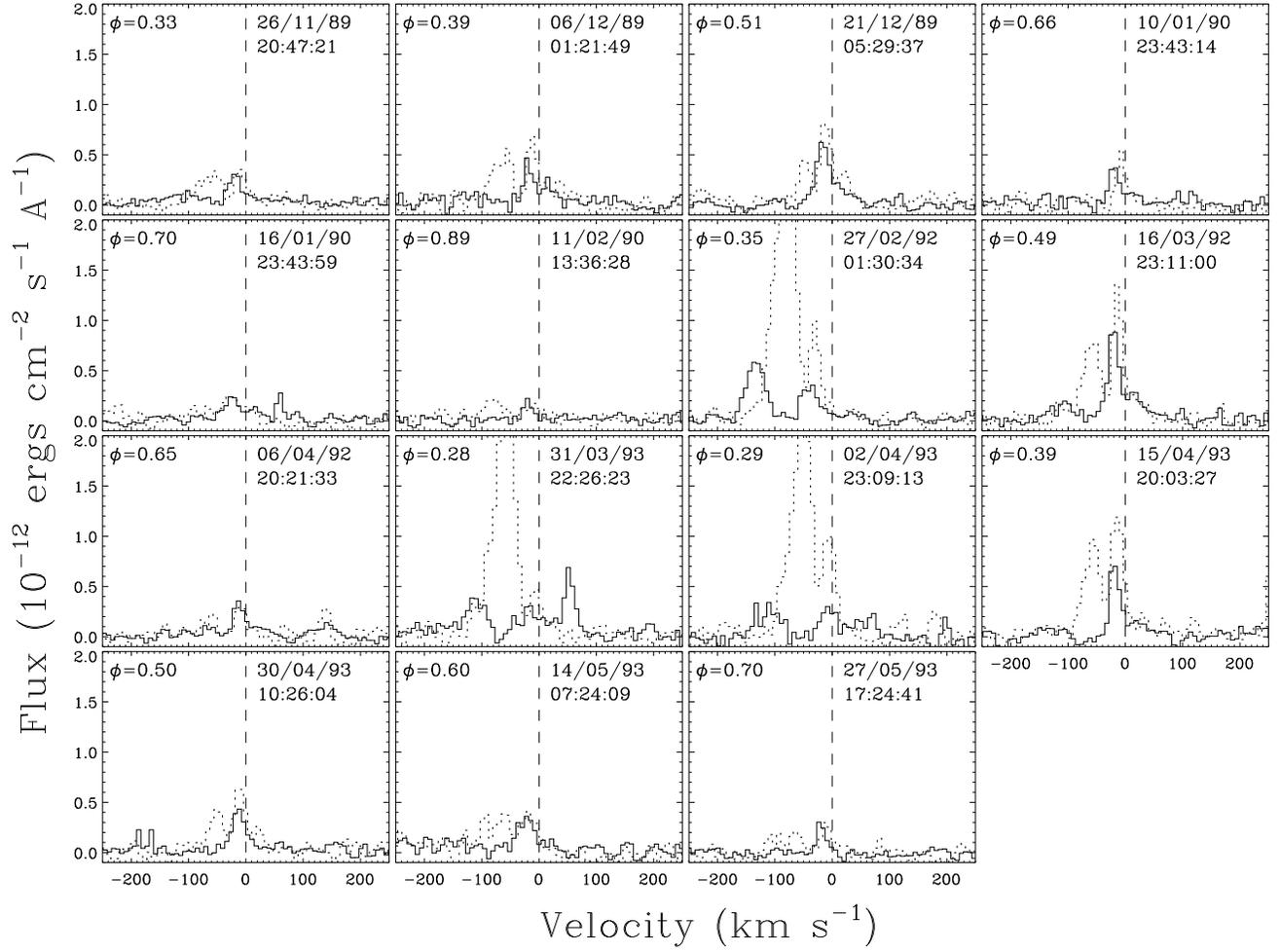}{3.5in}{90}{75}{75}{300}{0}
\caption{Same as Fig.\ 3, for L$^{2}$~Pup.}
\end{figure}

\clearpage

\begin{figure}
\plotfiddle{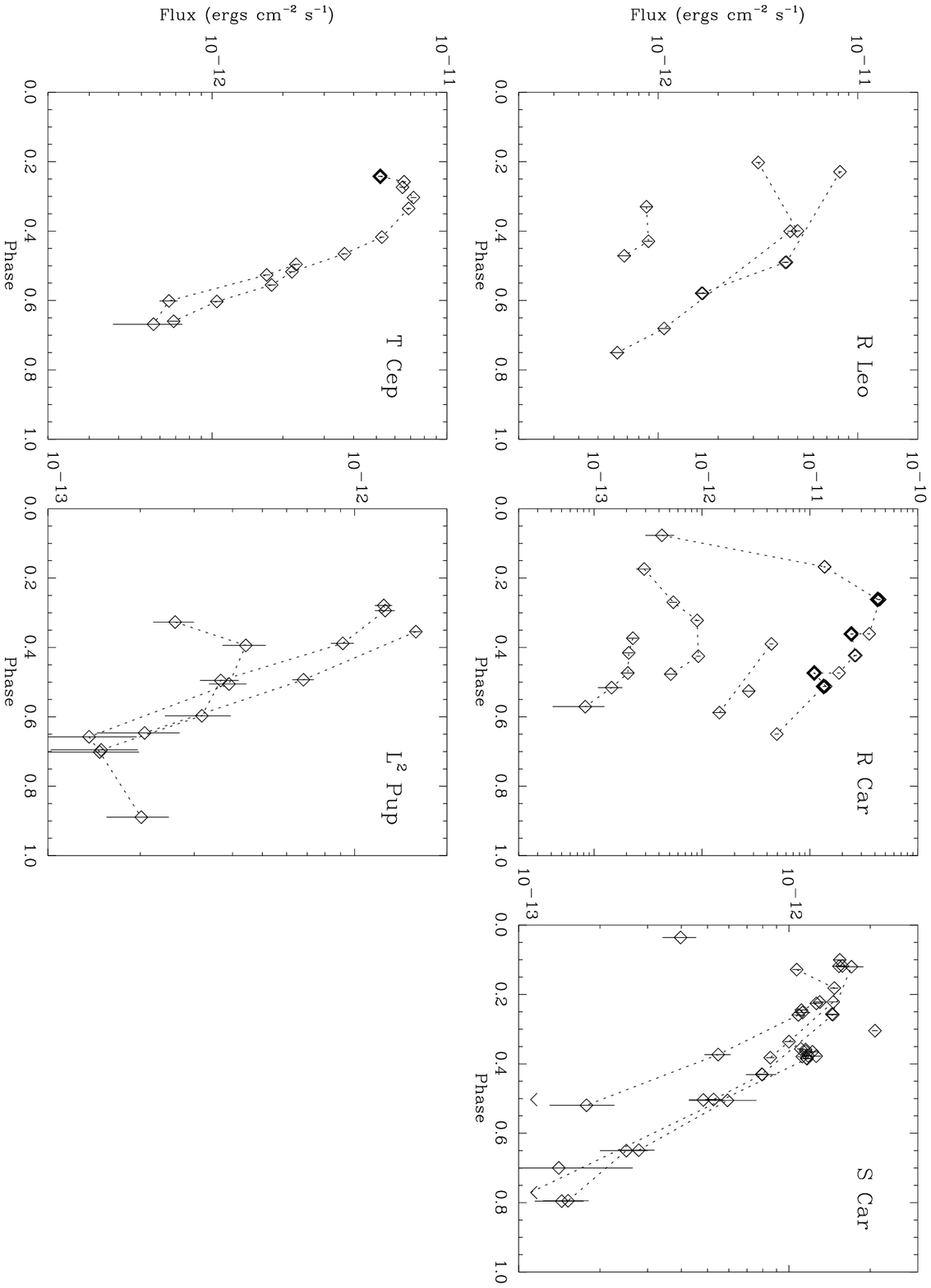}{3.5in}{90}{75}{75}{300}{0}
\caption{The Mg~II h line fluxes measured from IUE LW-HI spectra, plotted
  versus pulsation phase.  Dotted lines connect points within the same
  pulsation cycle.  Thick symbols identify potentially inaccurate data
  points due to overexposure, as indicated by NEWSIPS data quality flags.}
\end{figure}

\clearpage

\begin{figure}
\plotfiddle{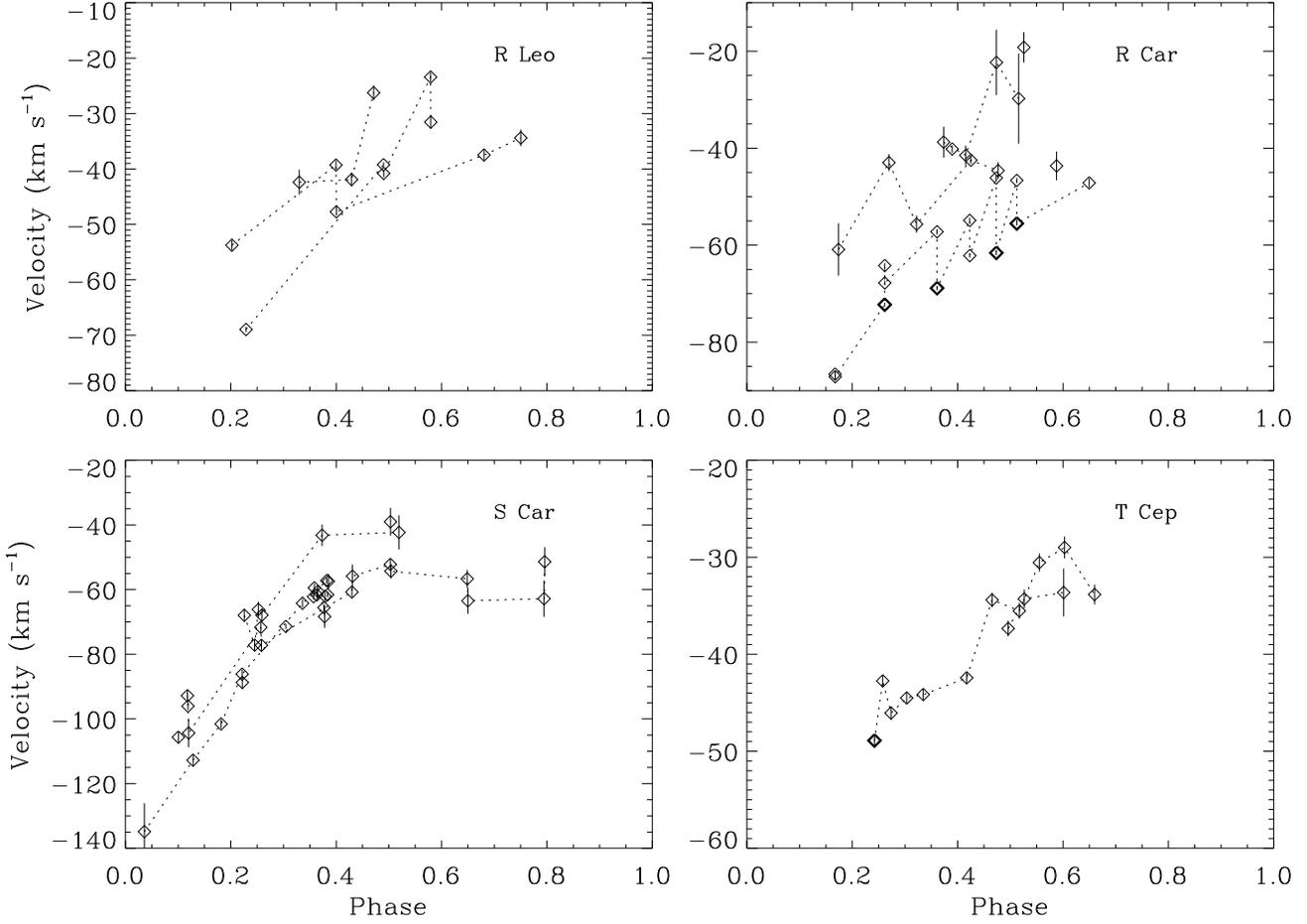}{3.5in}{90}{75}{75}{300}{0}
\caption{The Mg~II h line centroid velocities measured from IUE LW-HI spectra
  in the stellar rest frame, plotted versus pulsation phase.  Dotted lines
  connect points within the same pulsation cycle.  Thick symbols identify
  potentially inaccurate data points due to overexposure, as indicated by
  NEWSIPS data quality flags.}
\end{figure}

\clearpage

\begin{figure}
\plotfiddle{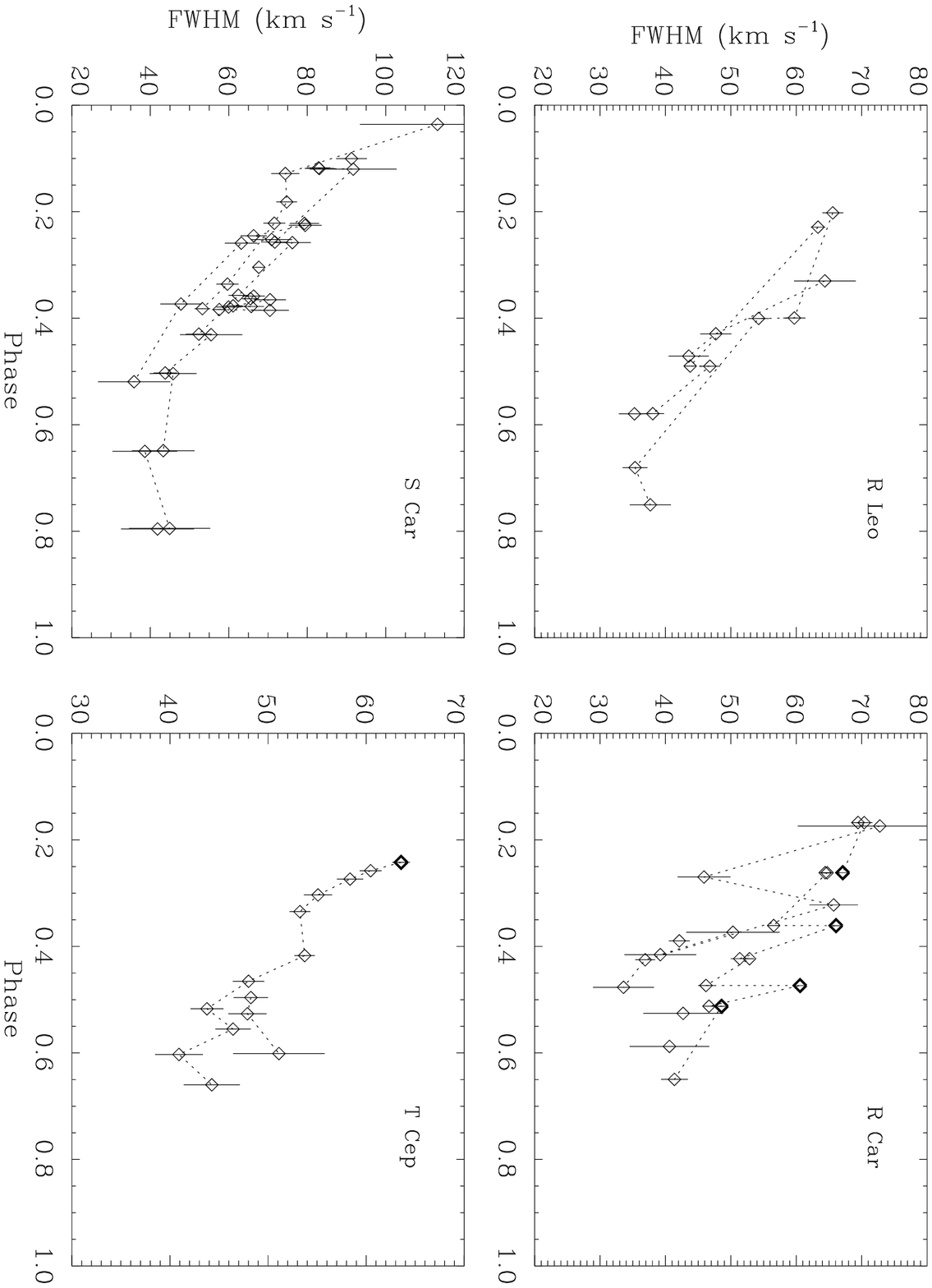}{3.5in}{90}{75}{75}{300}{0}
\caption{The Mg~II h line widths measured from IUE LW-HI spectra, plotted
  versus pulsation phase.  Dotted lines connect points within the same
  pulsation cycle.  Thick symbols identify potentially inaccurate data
  points due to overexposure, as indicated by NEWSIPS data quality flags.}
\end{figure}

\clearpage

\begin{figure}
\plotfiddle{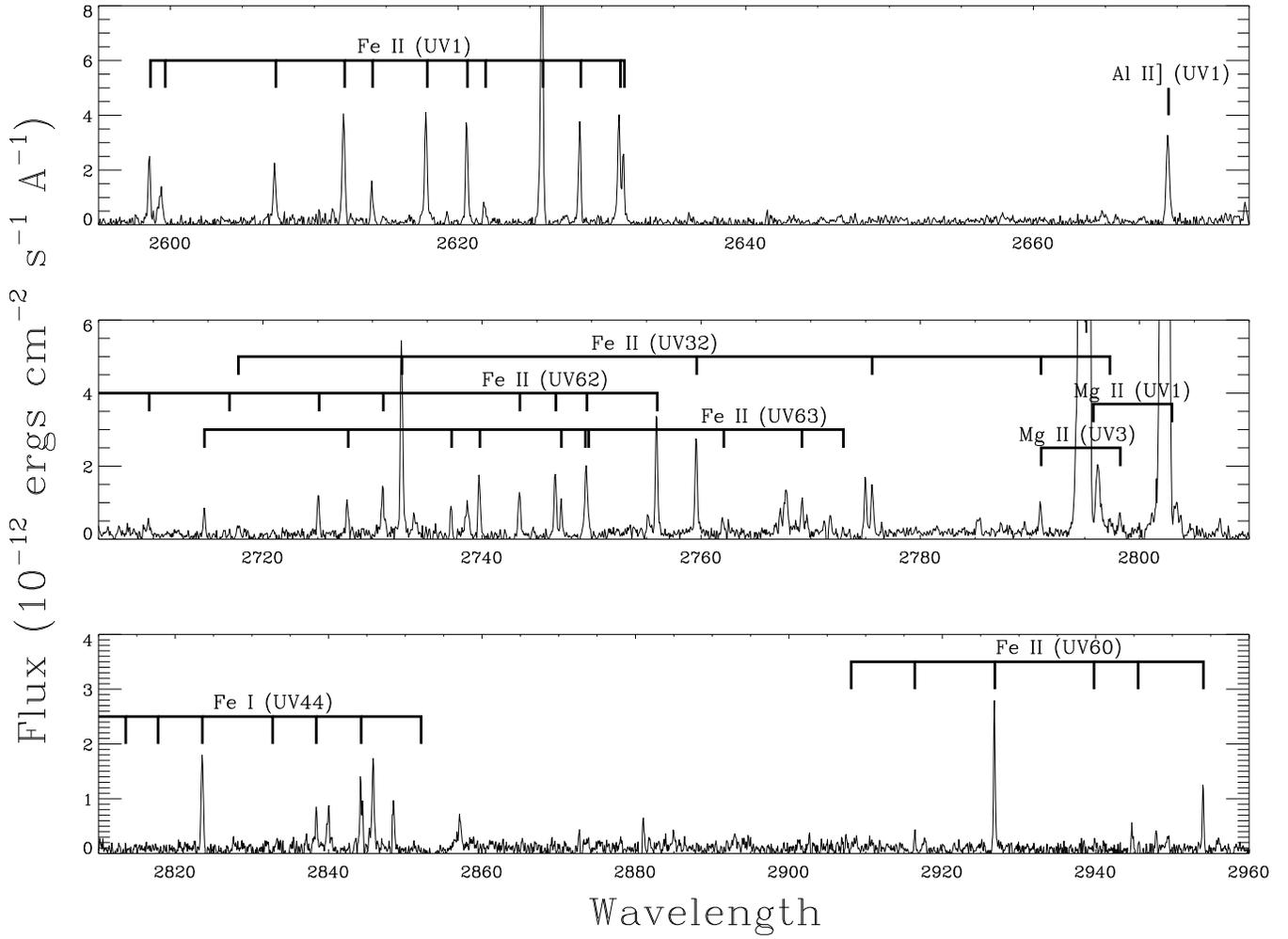}{3.5in}{90}{75}{75}{300}{0}
\caption{Sections of an IUE LW-HI spectrum (LWP17263) of R~Car obtained
  on 1990 January 30.  Expected line locations are displayed for several
  multiplets, which account for most of the lines seen in the spectrum.}
\end{figure}

\clearpage

\begin{figure}
\plotfiddle{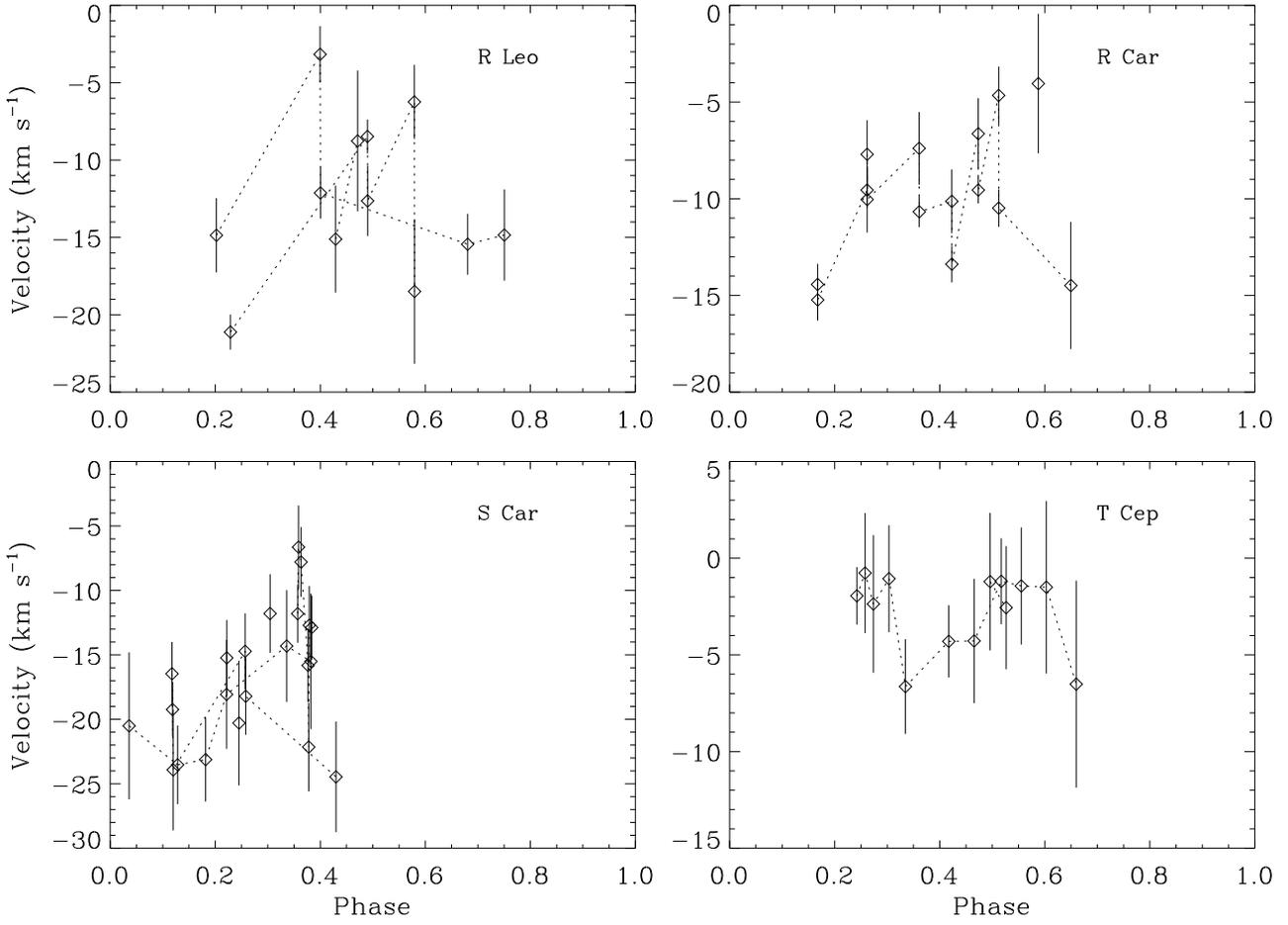}{3.5in}{90}{75}{75}{300}{0}
\caption{The centroid velocities of the Fe~II 2625.667~\AA\ line in the
  stellar rest frame, measured from IUE LW-HI spectra and plotted versus
  pulsation phase.  Dotted lines connect points within the same pulsation
  cycle.}
\end{figure}

\end{document}